\documentclass[%
 reprint,
 amsmath,amssymb,
 aps,
pra,
]{revtex4-2}

\usepackage{graphicx}% Include figure files
\usepackage{dcolumn}% Align table columns on decimal point
\usepackage{bm}% bold math
\usepackage{hyperref}% add hypertext capabilities
\usepackage{xcolor}
\usepackage{mathrsfs}
\newcommand{\psiq}{\Psi(\bm{q}_1,\bm{q}_2)}
\newcommand{\psix}{\psi(\bm{x}_1,\bm{x}_2)}

\begin{document}

\preprint{APS/123-QED}

\title{Optimizing the qudit dimensions of position-momentum entangled photons for QKD}

\author{Vikas S Bhat}
\author{Rounak Chatterjee}
\author{Kiran Bajar}
\author{Sushil Mujumdar}
% \author{Weakass S Butt}
% \author{Rawnuts Chatterbatterjee}
% \author{Carrom Gajar}
% \author{Sushi Mazedar}
    \email{mujumdar@tifr.res.in;http://www.tifr.res.in/~nomol}
\affiliation{%
      Tata Institute of Fundamental Research, 400005 Mumbai, India
    }%

\date{\today}% It is always \today, today,

\begin{abstract}
    We propose an optimization scheme to maximize the secure key rate of a high-dimensional variant of BBM92. We use the position-momentum conjugate bases to encode the higher dimensional qudits, realised in a fully passive optical setup.  The setup employs a single lens for the basis measurements and no lossy or slow elements. We optimize the qudit dimension for the protocol by maximizing the number of equiprobable sections (macropixels) of the detected beam while minimizing their overlap error. We show the enhanced key rate by discarding events from the ambiguous border pixels. Our strategy maximizes the overlap between the discarded regions from neighbouring macropixels, thereby globally minimizing the overall loss and error. We calculate the optimal dimension and the secure key rate for certain beam parameters. We experimentally show the feasibility of our scheme. This work paves the way for realistic implementations of high-dimensional device-independent quantum key distribution with enhanced bitrates.
\end{abstract}

\maketitle

\section{Introduction\label{sec:introduction}}
   In today's information age, the exponential growth of global data transmission \cite{cisco,youtube} has been coupled with an even faster growth in the security breaches of data \cite{cisco}. These data breaches are not just a concern for cyber-security experts but for the general public as well. With error-corrected quantum computing now being implemented on more and more qubits \cite{google, trapped_atoms, trapped_atoms_too}, our current encryption schemes are increasingly vulnerable to Shor's algorithm \cite{shor}. An inevitable alternative is information-theoretic secure encryption guaranteed by the laws of quantum mechanics, which have led to a few protocols. For instance, a quantum key distribution (QKD) protocol proposed by Brassard and Bennet in 1984 called BB84 \cite{bb84} encodes a random binary key in the polarization of single photons. Once distributed, the two parties share a string of identical random bits that can be used as a key for secure communications. Similarly, in 1991, Artur Ekert introduced the E91 protocol \cite{e91}, which utilized quantum entanglement to enhance further security, after which a variant of BB84 using entangled photons called BBM92 \cite{bbm92} was proposed. Although attractive for their unparalleled security, the actual implementation of these protocols using photon polarization is limited to a restrictive key generation rate of up to 0.5 bits/photon. This hinders the technology's widespread adaptability. 
   
   The widely accepted remedy towards this drawback is high-dimensional QKD (HDQKD) where the keys are encoded in higher dimensions. The benefit of implementing HDQKD is two-fold. First, the enhanced dimension size increases the bitrate and second, the inherent security and robustness to noise are enhanced \cite{enhanced_security, noise_robust,noise-reselience}. For example, the generalization of BB84 for $d$ dimensions leads to a bitrate of $\log_2{d/2}$ per qubit and an error threshold for an optimal cloning attack of $(\sqrt{d}-1)/2\sqrt{d}$ \cite{enhanced_security}. These benefits have led to realizing various candidates for HDQKD using orbital angular momentum of entangled photons \cite{oam1,oam2,oam3,oam-referee,oam-intracity}, time-bin qudits \cite{time-bin0,time_bin1,time_bin2,time_bin3,time-bin,time_bin4,time_bin5}, transverse momentum \cite{xp1,xp2,xp3,xp4,Lib-mplc-transverse} and polarization-momentum-time hybrid degrees of freedom \cite{hybrid1,sdm,hybrid2,hybrid-stack}. These implementations, however, also come with certain demerits. The time-bin implementations suffer from high errors when scaled due to phase instability. The rest require specialized optical elements like a spatial-light modulator, a raster scanning pinhole, a digital micromirror device, a multicore fibre etc, that are either slow, lossy or hard to scale. Effectively, either the qudit rate or the dimension is constrained. 

    In this work, we maximize the key rate of a completely optically passive implementation of a position-momentum higher dimensional variant of the BBM92. The conjugate bases of position and momentum are easily realised using a single optical lens via imaging or Fourier transform. We use the underlying structure of the wavefunction of entangled photons generated by spontaneous parametric downconversion (SPDC) to bifurcate the transverse beam's detection events into $d$ independent and equiprobable segments. To form these segments, we group pixels on the detector into macropixels. We find the optimal segmentation by maximizing the total number of macropixels while simultaneously minimizing the error due to crosstalk between them. We also show that, counter-intuitively, discarding certain events leads to a gain in the secure key rate. We use this strategy to calculate the optimal segmentation for a set of wavefunction parameters that are available to us. Experimentally, we realise a secure key rate of $3.2\pm0.1$ bits per photon. Our method is only limited by the generation rate of entangled photons, the detection rate, and the dark counts, promising significantly higher bit rates. 

    The rest of the paper is divided into four sections and two appendices. In Sec.~\ref{sec:wavefunction} we discuss the properties of the SPDC wavefunction and how to use it for HDQKD. In Sec.~\ref{sec:optimizing_qdit} we motivate our segmentation strategy that maximizes the secure key rate and show how to perform the segmentation for any given SPDC wavefunction. We present our experimental method and results in Sec.~\ref{sec:methods} and the conclusions in Sec.~\ref{sec:conclusion}. Finally, the two appendices provide mathematical details about our strategy's advantages.

\section{The SPDC wavefunction\label{sec:wavefunction}}
\begin{figure}
        \centering
        \includegraphics[width=0.6\linewidth,trim= 0.2cm 0 0.2cm 0,clip]{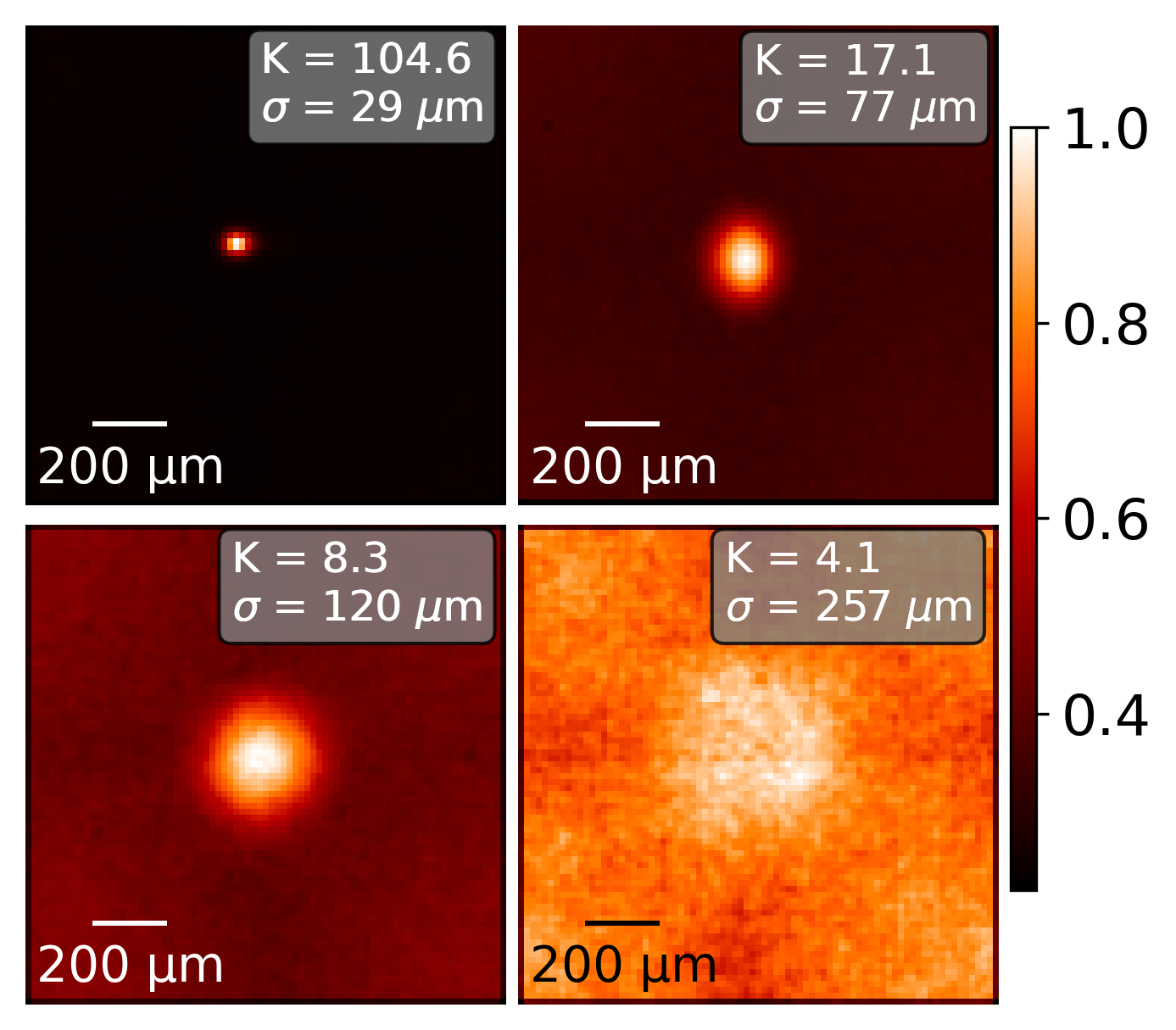}
        \caption{Experimentally determined convolutions in $(q_{x_1}+q_{x_2})$ vs $(q_{y_1}+q_{y_2})$ space for various Schmidt numbers depict the conditional distribution of the second photon, given the first photon was detected at the centre. The distribution follows a Gaussian of width $\sigma_{2|1}$. The increasing $\sigma_{2|1}$ with decreasing $K$ represents the increasing uncertainties in the conditional distribution of the partner photon. The $w_0$s used are $150,62,43,30$ microns.}
        \label{fig:convolutionVSwaist}
    \end{figure}

    We first describe the properties of the double-Gaussian wavefunction that can be used for HDQKD. When a Gaussian pump beam is incident onto a down-converting nonlinear crystal, the wavefunction describing collinearly down-converted SPDC photons at the crystal’s centre can, under well-justified approximations \cite{birthzone,oemz}, be represented as \cite{law&eberly, birthzone}
    \begin{eqnarray}
        \Psi\left(\boldsymbol{q}_1, \boldsymbol{q}_2\right)\propto e^{-\frac{w_0^2\left|\boldsymbol{q}_1+\boldsymbol{q}_2\right|^2}{4}} e^{-\frac{b^2\left|\boldsymbol{q}_1-\boldsymbol{q}_2\right|^2}{4}},\label{eq:dg_mom}\\
        \psi\left(\boldsymbol{x}_1, \boldsymbol{x}_2\right)\propto e^{-\frac{\left|\boldsymbol{x}_1+\boldsymbol{x}_2\right|^2}{4w_0^2}} e^{-\frac{\left|\boldsymbol{x}_1-\boldsymbol{x}_2\right|^2}{4b^2}},\label{eq:dg_pos}
    \end{eqnarray}
    where \(\boldsymbol{q}_i = (k_{ix}, k_{iy})\) represents the transverse momentum, \(\boldsymbol{x}_i = (x_i, y_i)\) denotes the transverse position of the \(i^{\text{th}}\) photon, \(w_0\) is the pump beam waist, and \(b^2 = L/3k_p\), with \(L\) as the crystal length and \(k_p\) the pump’s wavevector magnitude. This particular form of the wavefunction called the Double-Gaussian approximation (DG), is well-known for its accurate predictions of conditional and marginal statistics of biphotons\cite{birthzone}. Equation \eqref{eq:dg_mom} describes the state in the transverse momentum basis, while Eq. \eqref{eq:dg_pos} in the transverse position basis. In both forms, the first exponential term captures the Gaussian nature of the pump, and the second exponential arises from approximating the phase-matching function \cite{birthzone}. When $w_0 \gg b$, the first Gaussian in $\psiq$ forces $\bm{q}_1\approx-\bm{q}_2$ while the second Gaussian in $\psix$ enforces $\bm{x}_1\approx\bm{x}_2$. Thus, if both the photons are projected onto the position (momentum) basis, the outcome is correlated (anti-correlated). But, if they are projected onto different bases, there is no correlation in their outcomes. In particular, the uncertainty about $\bm{q}_1\approx-\bm{q}_2$ is given by $1/w_0$ while for $\bm{x}_1\approx\bm{x}_2$, it's given by $b$. We can measure these uncertainties using an EMCCD by the auto-correlate/auto-convolve method described in Ref.~\cite{edgar}. Figure~\ref{fig:convolutionVSwaist} illustrates the uncertainty $\sigma_{2|1}$ in the $(q_{x_1}+q_{x_2})$ vs $(q_{y_1}+q_{y_2})$ space for various $w_0$ computed via a background subtracted auto-convolution on the (30k) images captured by EMCCD. One of the measures to capture these uncertainties is the \textit{average} Schmidt number ($K$), which quantifies spatial entanglement. As seen in Fig.~\ref{fig:convolutionVSwaist}, the higher the $K$, the lower the uncertainty $\sigma_{2|1}$. For the DG, it has a closed form of \cite{law&eberly, oemz}
    \begin{eqnarray}
        K = \frac{1}{4}\left(\frac{b}{w_0}+\frac{w_0}{b}\right)^2.\label{eq:schmidt_number}
    \end{eqnarray}

    In principle, we have $K$ number of orthogonal modes that can be encoded for the HDQKD scheme. These are the modes resulting from the Schmidt decomposition, which, for the DG are the Hermite-Gauss (HG) or Laguerre-Gauss (LG) modes. Interestingly, HG modes and LG modes are mutually partially unbiased which has inspired a novel HDQKD protocol \cite{mpub}. The latter modes have been employed to realise OAM-based HDQKD \cite{oam1,oam2,oam3}. Although such modes naturally exist within the DG, they are not, a priori, equiprobable. Using them for HDQKD requires special preparation methods \cite{torres_oam, padgett_oam}. Furthermore, an xy-translated Gaussian mode packs more information when compared to OAM modes \cite{oam_bad}. However, the position-momentum entanglement present in the SPDC photons can be readily accessed with a single passive element, a lens. Along with a pixelated single photon detector, this lens acts as an all-in-one projective measurement device in the Cartesian coordinates for position when imaging the crystal, or momentum when placed in the Fourier plane of the crystal. 

\section{Optimizing the encoding dimensions}\label{sec:optimizing_qdit}
    The position-momentum based HDQKD protocol we optimize for is as follows. The sender (Alice) generates an entangled pair of DG photons, keeps one photon for herself and sends the partner to Bob. Alice and Bob independently choose to measure either the photon's position or the momentum on a pixelated detector. If both choose to measure the position (momentum) then their outcome would be correlated (anti-correlated). Thus, if the pixels of the detector encode the qudit information, then, after discarding the instances when the basis had mismatched, Alice and Bob should share an identical random secret key based on the string of activated pixels. In such a protocol, the way we segment the detector plane into macropixels plays a critical role in determining the efficiency, fidelity and ultimately the secure key rate. The aim is to find the best segmentation such that each macropixel activation is equiprobable while minimizing the cross-talk between them. The former condition ensures that no information is leaked via qudit-asymmetry and the latter minimizes the quantum dit error ratio (QDER). In this section, we show a segmentation strategy that significantly reduces cross-talk and increases usable information per sent photon. 

    \subsection{The benefit of discarding some pixels}\label{subsec:discard-rationale}
\begin{figure}
            \centering
            \includegraphics[width=0.9\linewidth,trim= 0.4cm 0.75cm 0.2cm 0.8cm,clip]{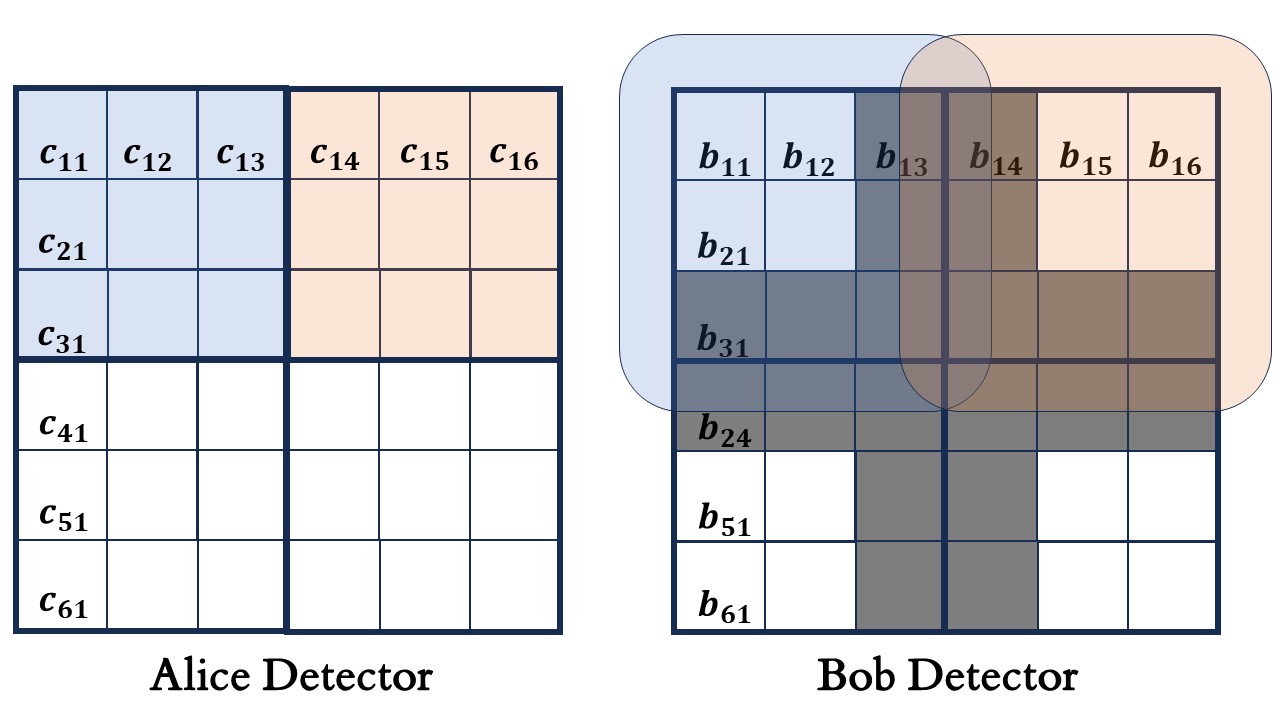}
            \caption{The two $6\times6$ pixelated detectors are divided into $4$ equal macropixels (indicated by the thicker edges). The blue and orange shading in the Alice detector corresponds to two mutually exclusive detection events. These colour-coded events correspond to partner detections on Bob's detectors. Since $2\sigma_{2|1}\approx1$ pixel, some of events in Alice's $c_{14}$ end up in Bob's $b_{13}$. The grey pixels on Bob's detectors correspond to the $20$ border pixels. Discarding events from the border pixels yields a higher secure key rate than when we don't.}
            \label{fig:discard_benefit}
        \end{figure}

        The secure key rate for a QKD protocol with $d$ dimensions of encoding and an error of $\varepsilon$ is given by \cite{secure-key-rate}:
        \begin{eqnarray}\label{eq:secure-rate}
            \mathcal{R}(d,\varepsilon)&=&\log_2d \nonumber\\&+& 2\varepsilon\log_2\left(\frac{\varepsilon}{d-1}\right) + 2(1-\varepsilon)\log_2(1-\varepsilon),
        \end{eqnarray}
        where the last two terms represent the negative high dimensional variant of Shannon entropy. Here, we present an observation. The secure key rate strongly depends on the error. In the protocol to be optimized, there is an intrinsic error due to the cross-talk between the macropixels at the shared borders. There is a disproportional gain in secure key rate from discarding these border pixels. For example, consider the case when Alice and Bob measure the photon's position using a detector with $6\times6$ pixels divided into $d=4$ equal macropixels, depicted in Fig.~\ref{fig:discard_benefit}. Let the Schmidt number be such that $2\sigma_{2|1}$ is about a pixel. For simplicity, we assume the photon detection is uniform over all the $36$ pixels. This assumption, while for illustrative purposes, contains the crux of the strategy that holds for the DG as well. Now examine Bob's pixel $b_{13}$. If activated, it could have originated from an Alice event at $c_{13}$, $c_{23}$, $c_{14}$, or $c_{24}$, where the last two possibilities lead to erroneous macropixel-tagging. There will always be such errors due to ambiguity at border pixels where a border pixel is any pixel adjacent to a different macropixel. In our example, there are $20$ border pixels, shown in grey in Fig.~\ref{fig:discard_benefit}. In this scenario, the error is $\varepsilon\approx0.16$ (see Appedix~\ref{ap:border-error} for the calculation). Using Eq.\eqref{eq:secure-rate} we get $\mathcal{R}(4,0.16)\approx0.22$ bits per photon. 
        
        Since all the error is coming from the border pixels, what if we discard all these events? This would lead to $\varepsilon=0$ but, as we now only detect $p=16/36$ of the events, our rate needs to be modified to
        \begin{equation}\label{eq:mod-keyrate}
            \mathcal{R}_\text{mod}(d,\varepsilon,p) = p~\mathcal{R}(d,\varepsilon).
        \end{equation}
        We propose this modified key rate $\mathcal{R}_\text{mod}$ as the relevant parameter to maximize. It factors in the fraction of photons detected by Bob that contribute to the final secret key and can be directly translated to the time required to generate the key. Using this formula, we get the final secure key rate of $\mathcal{R}_\text{mod}=0.89$ bits per photon. This is more than quadruple the earlier rate ($0.22$ bits per photon) we got without discarding events. Note that this enhanced rate is despite forgoing more than half of the events! Thus, we see a benefit in discarding events from the border pixels. 

        On the other hand, if our macropixel was $1$ pixel big, the error would be $\varepsilon\approx0.59$, and there would be $27$ border pixels leaving $d=9$ and $p=9/36$. In this case, the non-discarded rate is $R(9,0.5)<0$ and the modified rate $\mathcal{R}_\text{mod}(9,0,9/36)=0.79$ bits per photon which is lesser than what we achieved with $4$ macropixels. Thus, there is a trade-off: maximizing $d$ increases information per photon, but it also increases cross-talk. Naively maximizing $d$ is a suboptimal strategy to maximize the key rate.

    \subsection{Segmentation strategy}\label{subsec:segmentation}
        We now present the optimization strategy for the DG photons. The aim is to divide the transverse downconverted beam detection events into equiprobable macropixels while minimizing the cross-talk between them. The beam structure is described by the marginal distribution calculated from the DG wavefunction as
        \begin{eqnarray}
            \rho(x_1,y_1)&=&\iint _{-\infty}^{+\infty}\left|\psi(x_1,y_1;x_2,y_2)\right|^2dx_2dy_2\nonumber\\
            &=&\mathcal{N}e^{-r^2/2\sigma^2},\label{eq:marginal_beam}
        \end{eqnarray}
        when expressed in polar coordinates. Here, $\sigma^2=(w_0^2 + b^2)/4$, $r^2=x_1^2+y_1^2$ and $\mathcal{N}=1/2\pi\sigma^2$. Equation \eqref{eq:marginal_beam} describes the distribution of detection events.
    
\begin{figure}[t]
        \centering
        \includegraphics[width=\linewidth,trim= 0 0.25cm 0cm 0,clip]{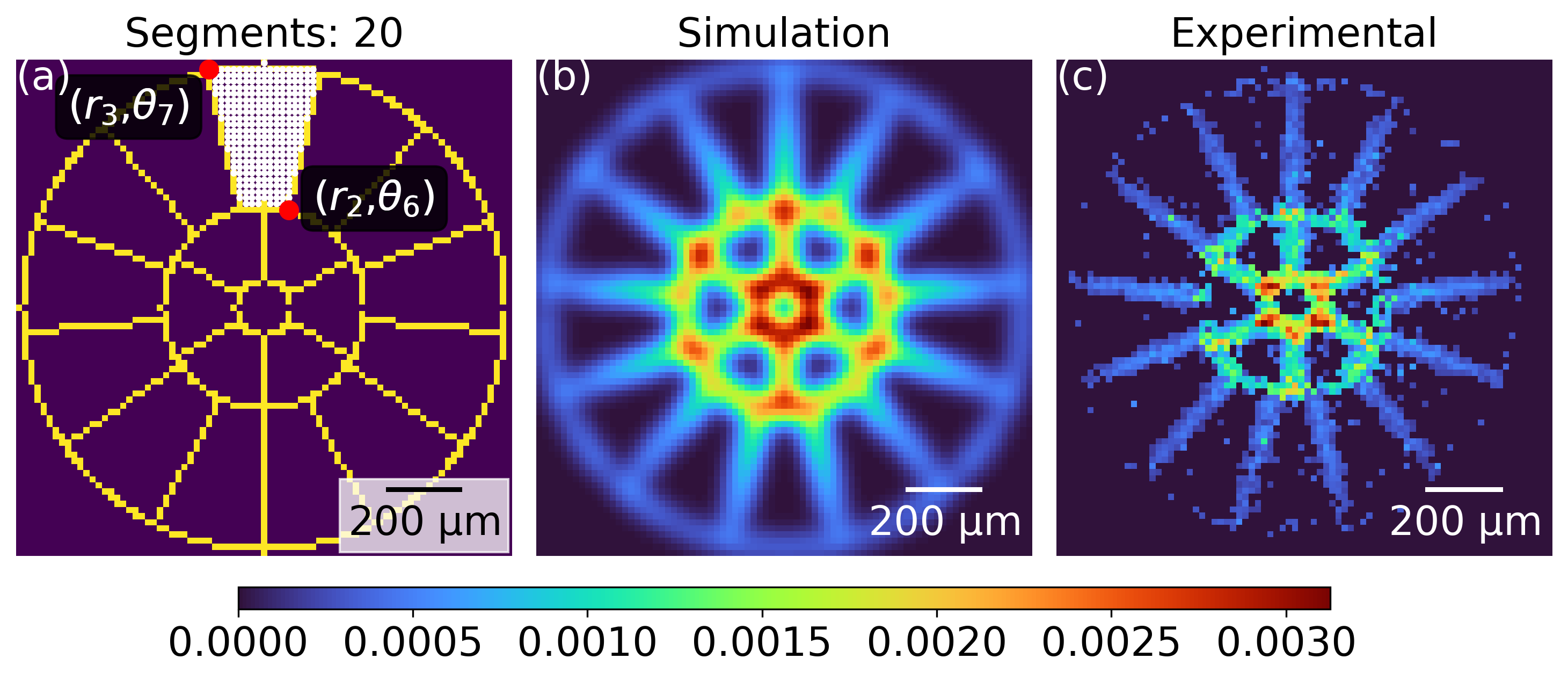}
        \caption{(a) The fryum wheel segmentation of the detector area into 20 macropixels (for $K=104.6$) with $R=(0,80,272,660)$ and $A=(1,6,13)$. One such macropixel is highlighted in white. Alice's border pixels are shown in yellow. The next two sub-figures depict the (b) simulated and (c) experimental photon distribution of Bob's momentum measurements when Alice's momentum measurements detected photons in the border pixels. These detection regions are ambiguous and need to be discarded. The outermost ring is shown only for illustrative purposes of the finite aperture and does not contribute to error calculation. The colourbar is in arbitrary units.}
        \label{fig:sectors}
    \end{figure}

        A prima-facie partitioning strategy would be to form $K$ number of close-packed hexagonal macropixels. However, this method is suboptimal as the DG form of the wavefunction implies that the hexagons away from the centre need to be larger to maintain equiprobability (See Appendix~\ref{ap:tiling} for more details). Next, the azimuthal symmetry of the wavefunction may suggest a division of the transverse beam into sectors of a circle. However, this method is also erroneous due to the following reasons. All the sectors converge at the centre with a high density of border pixels enhancing the QDER, a situation worsened by the centre being the most probable detection event. Therefore, we devise an optimum strategy that divides the detection area into $N$ radial segments $R:=(r_0=0,r_1,\dots,r_N=r_\text{ap})$ and azimuthal segments $A:=(a_0=1,a_1,a_2\dots,a_{N-1})$ where $r_i$ is the radius of the $i$th segment and $a_k$ is the number of macropixels lying between $r_{k}$ and $r_{k+1}$. The aperture of the relevant optics is $r_\text{ap}$. For example, in Fig.~\ref{fig:sectors}, we have $R:=(0,80,272,660)$ (in $\mu$m) and $A:=(1,6,13)$. To minimize the QDER, the innermost macropixel is always a circle ($a_0=1$). We call this the `fryum wheel' segmentation, named after the visually similar Indian snack.
        
        We now show how to calculate $R$ for a given $A$. Figure \ref{fig:sectors} illustrates the division of the detector into 20 macropixels as marked by the yellow boundaries. From Eq. \eqref{eq:marginal_beam}, the marginal density $\rho(r,\theta)$ provides the probability of finding a photon within a region defined by points $(r_i,\theta_i)$ and $(r_j,\theta_j)$ as
        
        \begin{eqnarray}\label{eq:seg_probability}
            P(r_i,\theta_i;r_j,\theta_j) = \mathcal{N}\int_{\theta_i}^{\theta_j}d\theta\int_{r_i}^{r_j}e^{-r^2/2\sigma^2}rdr.
        \end{eqnarray}
        In particular, we set the probability of finding a photon within any macropixel to $\alpha$. An example is shown in white in Fig.~\ref{fig:sectors}. The macropixels get bigger with their radial displacement to maintain equiprobability as the events decay radially. We define $c_k=\sum_{i=0}^{k-1}a_i$ as the total number of macropixels within the radius $r_k$. Now, using the fact that each macropixel needs to have a probability of $\alpha$, using Eq.~\eqref{eq:seg_probability} leads to the following condition on $r_k$
        \begin{eqnarray}\label{eq:radial_segs}
            &P(0,0;r_k,2\pi)=c_k\alpha = 2\pi\mathcal{N}\int_0^{r_k}e^{-r^2/2\sigma^2}rdr,\nonumber\\
            &\Rightarrow r_k = \sqrt{-2\sigma^2\ln(1-c_k\alpha)}.
        \end{eqnarray}
        Finally, to ensure that the final radial segment is always set to the optical aperture, $r_N=r_\text{ap}$, we add an auxiliary angular segment $a_\text{aux}$ for $r\in (r_\text{ap},\infty)$, found by setting $\alpha = 1/(c_N+a_\text{aux})$ (normalization condition) and inverting Eq.~\eqref{eq:radial_segs} as
        \begin{eqnarray}\label{eq:auxilary_seg}
            a_\text{aux}=c_N\left(\frac{1}{e^{r_\text{ap}^2/2\sigma^2}-1}\right).
        \end{eqnarray}

    \subsection{Finding the optimal segmentation}\label{subsec:optimal_segmentation}
        Having developed a method to construct the segmentation for any given list $A$, the goal now is to find the most optimal one. Our strategy to maximize the secure key rate is to, as described in Sec.~\ref{subsec:discard-rationale}, discard the overlapping region where the detection event is ambiguous. The pixels with ambiguous events are depicted in Fig.~\ref{fig:sectors} where, when Alice detects events on the yellow border pixels, Bob's detection events are spread about a width of $\sigma_{2|1}$. As seen from Fig.~\ref{fig:convolutionVSwaist}, when the sender's photon falls on a pixel, the conditional probability (posterior distribution) of the receiver's photon follows a Gaussian distribution of width $\sigma_{2|1}$. It is a well-known property of the Gaussian that its area beyond three times the standard deviation is less than $0.27\%$. Thus, if we discard pixels amounting to $3\sigma_{2|1}$  around the borders, the probability of Bob's macropixel detecting a neighbouring macropixel reduces to $<0.27\%$. This implies that the QDER (from cross-talk) should be reduced to $<0.27\%$. Since each segment needs to be equiprobable, the border pixels are discarded until each macropixel has the same probability. This corresponds to discarding a fraction of the events equalling $\mathcal{A_\text{discard}}$. As discussed in Sec.~\ref{subsec:discard-rationale}, there is, however, a trade-off. Discarding pixels from the total probability within the aperture ($c_N\alpha$) will lead to a reduction in the average bitrate. Thus, the constrained optimization to calculate the maximum secure key rate $\mathcal{R_\text{opt}}(d,\varepsilon,p)$ per photon with $d=c_N$ and $p=c_N\alpha-\mathcal{A_\text{discard}}$ becomes
        \begin{align}\label{eq:optimization_condition}
            \mathcal{R_\text{opt}}(d,\varepsilon,p)&=\text{Max}\big[\left(c_N\alpha - \mathcal{A}_\text{discard}\right)\times\mathcal{R}(d=c_N,\varepsilon)\big]_A,\nonumber\\
            &=\text{Max}\Bigg[\left(\frac{c_N}{c_N+a_\text{aux}}-\sum_{i\in\text{\{discard\}}}\rho(r_i,\theta_i)\right)
            \nonumber\\&~~~~~~~~~~~~~~~~\times\mathcal{R}(d=c_N,\varepsilon)\Bigg]_A,
        \end{align}
        where we maximize the product of the total secure key rate $\mathcal{R}$ and the total utilized probability (term in parenthesis). For example, the optimal segmentation for a DG with $K=104.6$ is found to be $A=(1,6,8,21,5)$ where $c_N=36$ and we got $p=0.66$ giving a secure key rate of $\mathcal{R}_\text{opt}=3.4$ bits per photon. Here the auxiliary angular segment is $5$ to set the maximum radius to fit within the aperture of our detector.

\section{Methods and Results}\label{sec:methods}
    \subsection{Experimental setup}
\begin{figure}[t]
        \centering
        \includegraphics[width=0.8\linewidth,trim=3cm 3.5cm 11.5cm 1cm, clip]{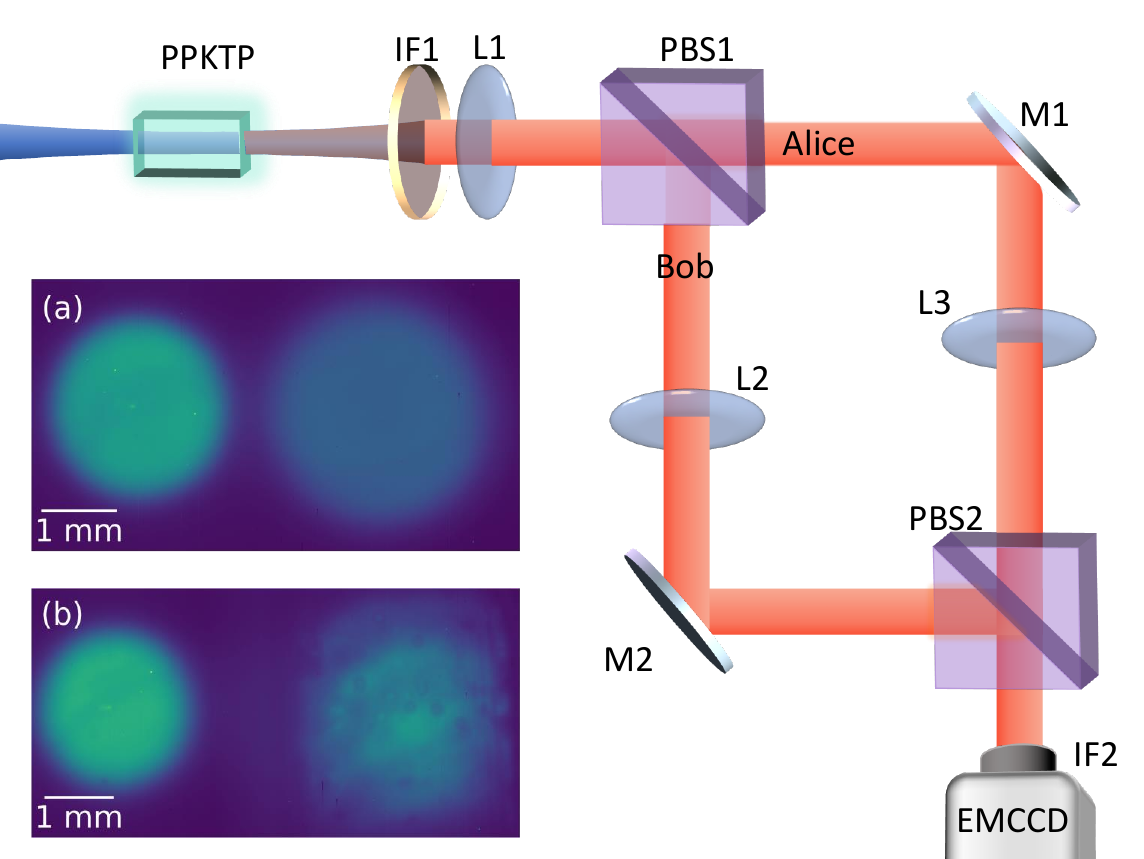}
        \caption{Experimental setup for high-dimensional BBM92. L$_\text{i}$: Lens, PPKTP: Periodically-Poled Potassium Titanyl Phosphate, IF: Interference Filter around 810~nm, PBS: Polarizing Beam Splitter, EMCCD: Electron Multiplying Charge Coupled Device, M: Mirror. The inset shows (a) momentum-momentum and (b) momentum-position beams on the detector.}
        \label{fig:setup}
    \end{figure}

        Here we describe the experimental setup used to verify the detector segmentation scheme described in Sec.~\ref{sec:optimizing_qdit}. The source of our spatially entangled photons, as seen in Fig.~\ref{fig:setup}, is a $10$mm long type-2 PPKTP crystal ($2\times1$mm$^2$ aperture) that is pumped by a narrowband laser at $405$ nm with $\sim 100\mu$W of power. The crystal is maintained at $40^\circ$C to ensure maximum degeneracy between the two photons. Interference filter IF1 blocks the excess pump. Lens L1 (focal length $f_1=35$ mm) is kept such that the crystal centre is in its front focal plane. A polarising beam splitter (PBS) deterministically separates the $H$ and $V$ polarizations, one of which is sent to Alice and the other to Bob. Alice uses L3 to measure momentum or position while Bob uses L2. A momentum-basis measurement is implemented by imaging the back focal plane of L1 onto the detector face using lens L2/L3 (focal length $f=75$ mm) in a 2f setup. A position basis measurement is implemented by using lens L2/L3 ($f=125$ mm) placed appropriately with respect to L1 to image the crystal centre on the detector. The two photons are then redirected to a second PBS which sends both the beams parallelly to the EMCCD.
        
        We acquired 3 million frames on the EMCCD at a temperature of $80^\circ$C, EM gain of $300$, pre-amp gain of $3$, horizontal readout rate of $5$MHz, vertical shift speed of $0.5$s and Vertical clock voltage amp of $+1$. We extracted the photon counts using a multi-threshold scheme described in Ref. \cite{rounak}. The pump power was set such that there was on average, $0.1$ photons per frame and its waist such that the Schmidt number $K=104.6$ throughout the experiment.

    \subsection{Extracting the secure key rate}
        Once we have the photon-counted frames, we first vertically divide each frame into an Alice frame $C_i^m$s and a Bob frame $B_i^m$s where $1\le i\le N_\text{frames}$ and $m=x$ for position measurement or $m=p$ for momentum measurement. Each $C_i$/$B_i$ is then subdivided into $d$ macropixels according to the fryum wheel segmentation $(A,R)$. The error matrix $\mathscr{E}_{ij}$ counts the probability of simultaneous detection events at Alice macropixel-$i$ and Bob macropixel-$j$. The counts detected in the $k$th Alice frame's $l$th macropixel in $m$ basis is denoted by $C_{k,l}^m$. The error matrix is then evaluated using
        \begin{align}\label{eq:error-matrix}
            \mathscr{E}_{ij}^{mm'}=\frac{1}{N_\text{frames}}\sum_{n=1}^{N\text{frames}} &C_{n,i}^m \times B_{n,j}^{m'} \nonumber\\
            &- C_{n,i}^m \times B_{n+1,j}^{m'},
        \end{align}
        where the subtracted term acts as a background correction to statistically remove the accidental coincidence. This works on the assumption that all partner photons fall in the same frame which is valid as the inter-frame time window ($\sim10$ms) is much greater than the photon pair coherence time, meaning all the inter-frame coincidences are accidental.

        \begin{figure}[t]
            \centering
            \includegraphics[width=.65\linewidth,trim= 0.35cm 0.2cm 0.25cm 0.2cm,clip]{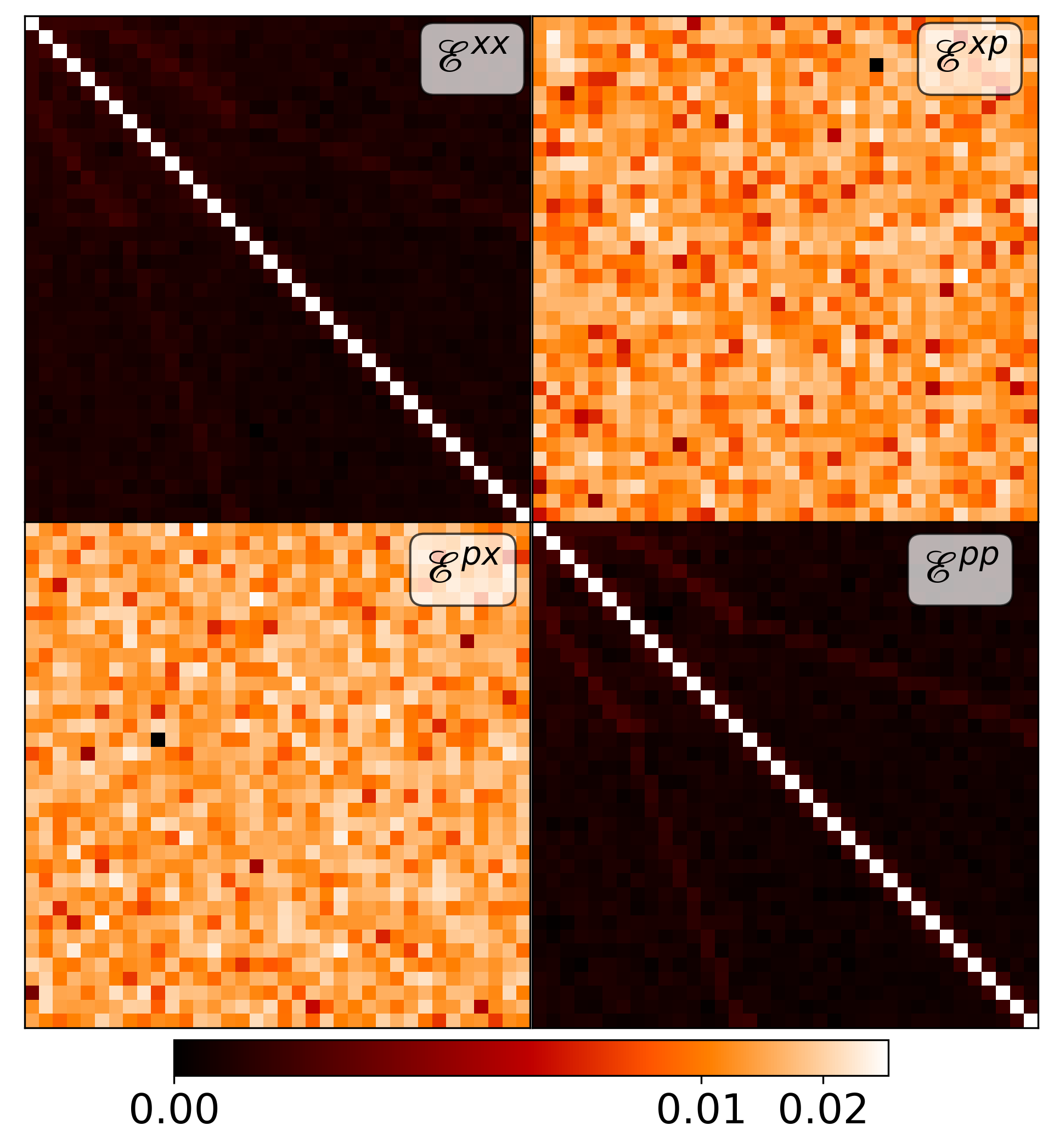}
            \caption{The measured $4\times(36\times36)$ error matrix $\mathscr{E}^{mm'}_{ij}$ with a $\gamma=0.4$ power-law scaling to show the miniscule off-diagonal terms. In clockwise from top-left we have the error matrix for position-position, position-momentum, momentum-position, momentum-momentum bases. The error estimated from this is $\varepsilon=0.98\%$ for the position basis and $\varepsilon=0.97\%$ for the momentum basis. This run led to a secure key rate of $3.2$ bits per photon.}
            \label{fig:error-matrix}
        \end{figure}

        The measured error matrix for an experimental run with $d=36$ is shown in Fig.~\ref{fig:error-matrix}. From this, we can estimate the QDER of the QKD protocol as follows,
        \begin{eqnarray}
            \varepsilon=1-\frac{1}{d}\text{Tr}(\mathscr{E}_{ij}^{mm'})=1-\frac{1}{d}\sum_{i,m}\mathscr{E}_{ii}^{mm}.
        \end{eqnarray}\label{eq:evaluated-qder}
        In the experimental run shown in Fig.~\ref{fig:error-matrix}, we were left with $\varepsilon=0.98\%$ for the position basis and $\varepsilon=0.97\%$ for the momentum basis. The remnant error is mostly from accidental coincidences that could not be corrected. To achieve these low errors, we had to discard $34.8\%$ of the events ($p=0.65$). Taking the error as $\varepsilon=0.98$, using Eq.~\eqref{eq:mod-keyrate} we estimate a secure key rate of $3.2$ bits per photon. In this run, the error without discarding the border pixels was $\varepsilon=28.2\%$, which leads to a secure key rate of only $0.6$ bits per photon.
        
        \subsection{Optimizing the hyperparameters of the fryum wheel}
            Finally, to experimentally validate our optimization strategy, we consider all possible valid segmentation configurations of length $N$. A segmentation is valid if:
            \begin{itemize}
            \itemsep0em
                   \item the first radial segment has one angular segment,
                   \item the radial segments are more than $3\sigma_{2|1}$ apart,
                   \item the angular segments are more than $3\sigma_{2|1}$ apart,
                   \item the last angular segment is set as auxiliary.
               \end{itemize}

            \begin{figure}[t]
            \centering
                \includegraphics[width=\linewidth,trim = 0 0 0 0,clip]{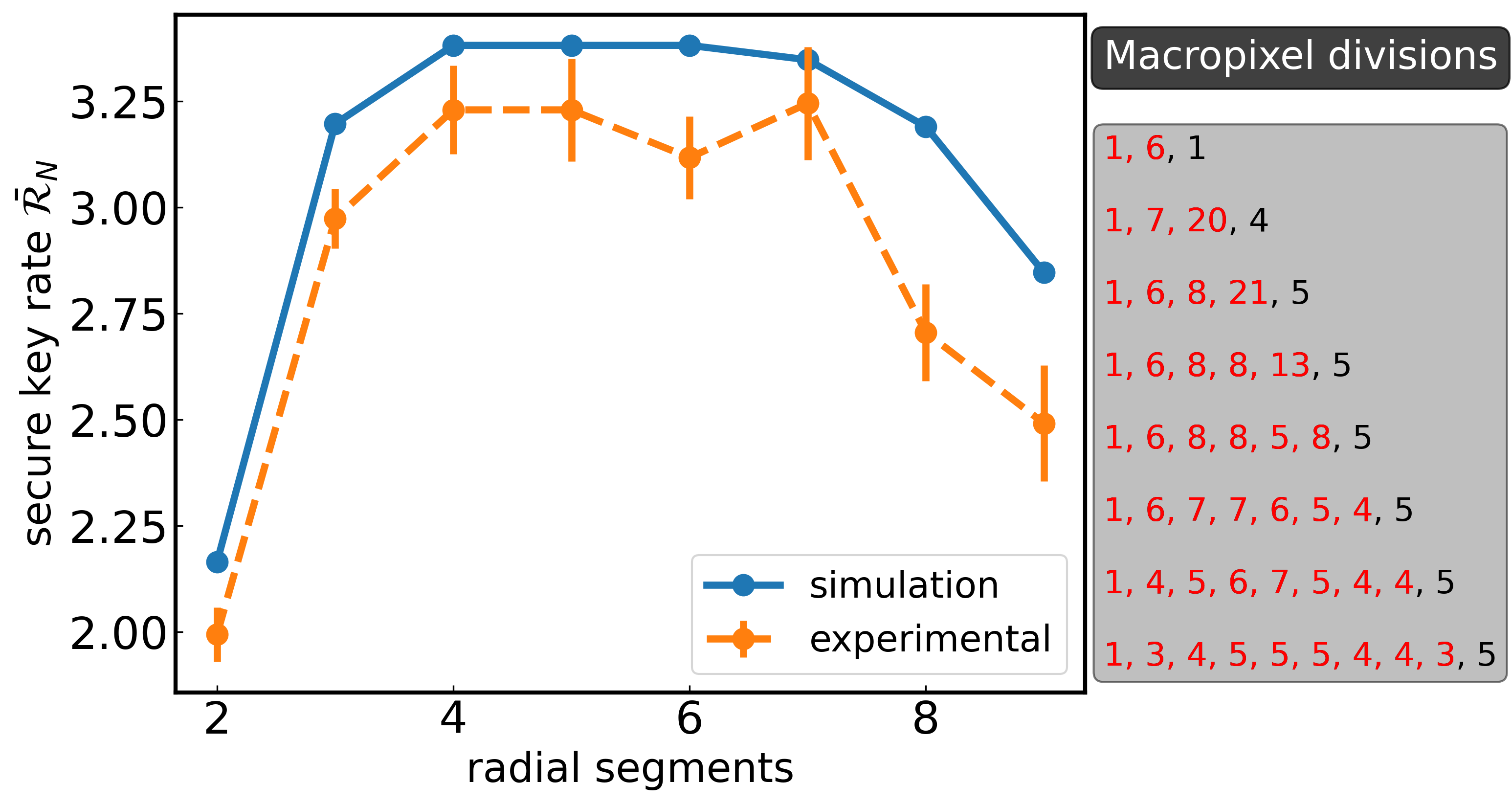}
                \caption{The result of our secure key rate ($\bar{\mathcal{R}}_N$) optimizer according to theory (blue) and experiment (orange with error bars) shows a clear maxima for both theory and experiment at $N=4$ radial segments with $A=(1,6,8,21,5)$. The box of corresponding macropixel divisions is shown on the right in grey. The black entries depict the auxiliary angular segment. These results are for $K=104.6$.}
                \label{fig:qber}
            \end{figure}
            
            Next, for each $i$th segmentation (of length $N$), we calculate the discarded area as described earlier, and from it, the secure key rate $\mathcal{R}_\text{opt}^{(i,N)}$. Their maximum is represented as $\bar{\mathcal{R}}_N\equiv\text{Max}[\mathcal{R}_\text{opt}^{(i,N)}]_i$. Repeat this for all possible $N$. Figure~\ref{fig:qber} shows $\bar{\mathcal{R}}_N$ vs $N$, the result of such an optimization for $K=104.6$ where $2\le N\le9$. There is a clear maximum at $N=4$ with $\bar{\mathcal{R}}_4=3.2\pm0.1$ bits per photon in our experimentally obtained rate (orange curve). The blue curve shows the theoretical prediction, which shows a maximum at $\bar{\mathcal{R}}_4=3.4$ bits per photon also peaks for the same macropixel segmentation of $A:=(1,6,8,21,5)$. The disagreement between the two curves can be attributed to the noise in our EMCCD. The noise increases when photons populate the pixels densely, a situation that holds for larger $N$. This noise arises primarily from low temporal resolution in EMCCDs, which can be alleviated by employing more sophisticated detectors like SPAD cameras \cite{spadcam1,spadcam3}. The SPAD cameras have a time tagging ability at a detection rate of $500$kHz with a dark count rate of $50$Hz. Since this is the limiting factor,we estimate our implementation can, in principle, produce a sifted photon rate of $\sim1$Mbits per second. 

            The rate can be further enhanced by increasing the pump waist size and decreasing the crystal length which enhances the $d$ of the system. We chose our parameters such that we achieved a $d=36$, as there exist multi-core fibers which support $36$ channels that have been tested for up to $2500$km\cite{36mcf1,36mcf2}. The rate can also be enhanced by multiplexing with a time-bin system \cite{hybrid-stack}.

\section{Conclusion}\label{sec:conclusion}
    In conclusion, we have demonstrated a method to maximize the secure key rate extracted from a higher dimensional variant of BBM92 using position-momentum entangled downconverted photons. We showed that the secure key rate is enhanced by discarding events from border pixels. Using the discard strategy, we devised an optimal segmentation scheme of the entangled photons, the fryum wheel model, for a given beam and crystal parameter by exploiting the underlying symmetry and structure of the wavefunction. Our experimental realization of this scheme resulted in a secure key rate of $3.2\pm0.1$ bits per photon, which shows a very good match with our theoretical predictions.
    
    Our segmentation can also be used to reverse-engineer the beam properties that maximize coupling into multicore fibres. This can enhance the bitrate in the space-division multiplexed fiber-based QKD \cite{sdm}. Our experimental parameters optimize for a $36$ channel multi-core fiber\cite{36mcf1,36mcf2}. Further, since efficient wavefront correction for entangled photons is well established \cite{yaron,kiran-immunity,mehul-unscramble,karimi-AO}, our method can be used in free-space position-momentum HDQKD for long distances as well. Thus, we expect our result will lead to newer technology implementing device-independent QKD with very high bitrates in both free-space and fibres.

\bibliographystyle{apsrev4-2}
\bibliography{apssamp}% Produces the bibliography via BibTeX.

\appendix

\section{Comparing segmentations for uniform probability distribution} \label{ap:tiling}
Here, we compare two more tiling strategies to our result. The task is to divide a circle of radius $R$ into as many smaller segments of area $\geq\pi r^2$, dubbed atoms. The additional constraint is that the borders of these segments must be a minimum $r$ distance apart.

\begin{figure}
    \centering
    \includegraphics[width=0.9\linewidth, trim= 0 0.3cm 0 0.25cm, clip]{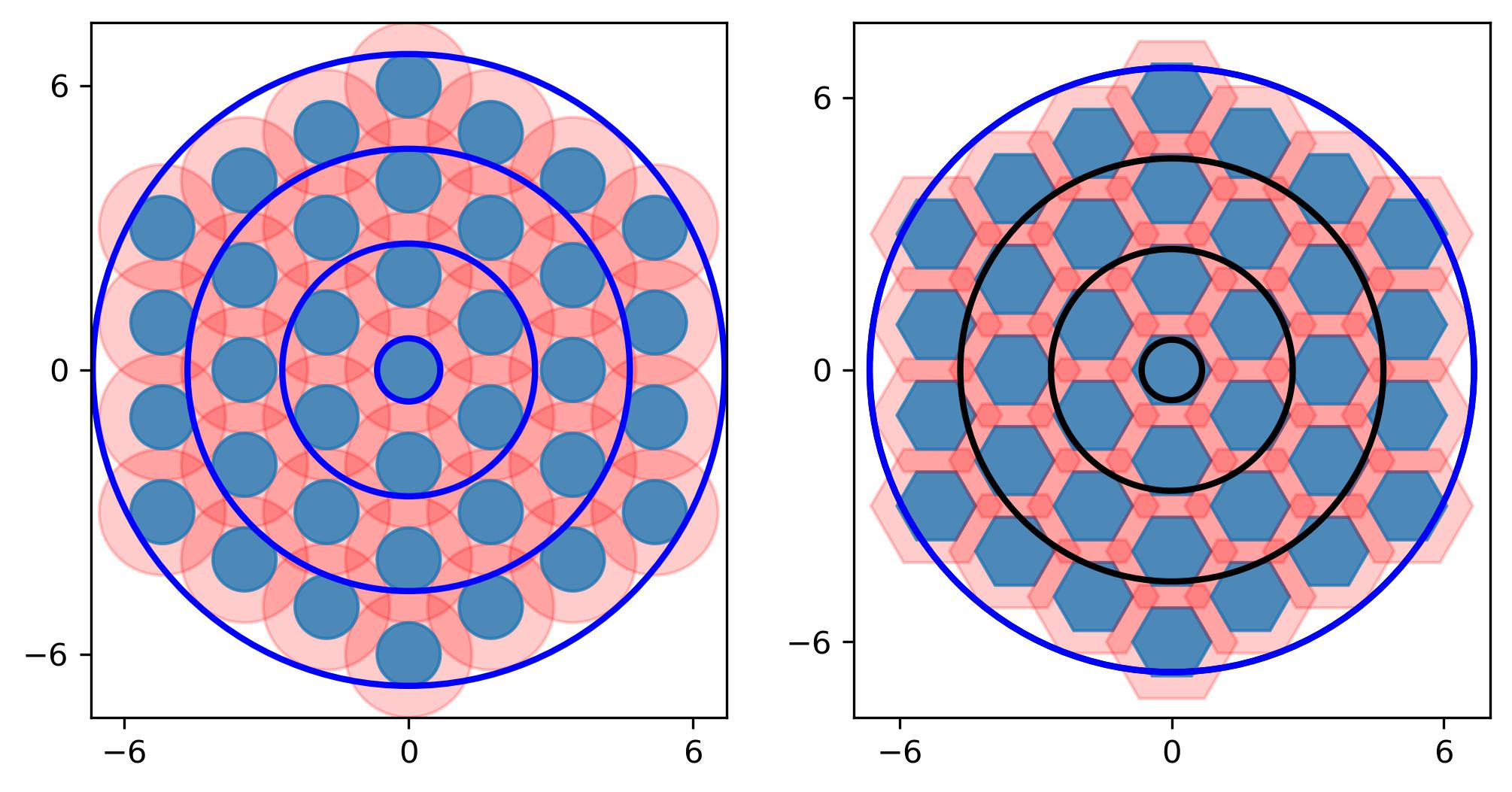}
    \caption{The hexagonal segmentation show circles (left) and hexagons (right) in each radial disk. The red translucent regions depict the discarded area. It can clearly be seen that the discarded regions do not overlap completely in the case of circles but do for hexagons.}
    \label{fig:packing}
\end{figure}

We first work out the optimal segmentation considering circular atoms of radius $r$. Since each circle is of radius $r$ and the separation between borders is also $r$, the distance between a circle and its neighbours is $3r$. That is, their centres form an equilateral triangular lattice of side $3r$ forming the lattice of a hexagonal packing structure. If we start populating from the centre of the outer circle, we get hexagonal packing patterns as shown in Fig.~\ref{fig:packing}(a). The solid blues are the atoms of radius $r$ and the translucent red region is the discarded region of width $r$. The radius of the $k$th radial segment is given by 
\begin{eqnarray}\label{eq:Rk}
    R_k=(3k-2)r,
\end{eqnarray} and we assume that the outer circle $R$ is such that $R=(3n-2)r$ where $n$ is the total number of radial segments (4 in Fig.~\ref{fig:packing}).

The number of atoms that fit in $R$ is then given by $N=3n^2-3n+1$. This gives the total discarded area as
\begin{eqnarray}
    A_\text{disc}^\text{circ} &=& \pi R^2 - N\pi r^2\nonumber\\
    &=& 6(n-1)\left(n-\frac{1}{2}\right)\pi r^2.
\end{eqnarray}

It can be bettered if we manage to maximally overlap the discarded regions between the neighbouring atoms. This is done by tiling with equilateral triangles, squares or hexagons. Hexagons, on virtue of being the smallest of these to contain a circle, offer the best option. The area of the smallest hexagonal atom containing a circle of radius $r$ is given by $2\sqrt{3}r^2$ with the side length $a=2r/\sqrt{3}$. See Fig.~\ref{fig:packing}(b). Thus, the total discarded area is
\begin{eqnarray}
    &A_\text{disc}^\text{hex}& = \pi R^2 - N \left(2\sqrt{3}r^2\right)\nonumber\\
    &=& \left(\left(9-\beta\right)n^2 - \left(12-\beta\right)n + 4 - \frac{\beta}{3}\right)\pi r^2,
\end{eqnarray}
where $\beta=6\sqrt{3}/\pi$. 

To ascertain that the fryum segmentation scheme is better, we need to show that we can fit in more segments than the hexagon scheme, while discarding lesser area. To that end, we divide the outer circle again into $n$ radial annuli of width $3r$, similar to the hexagon case. In each annulus, the inner annulus of width $r$ is discarded. The area of $k$th annulus for $k>1$, using Eq.~\eqref{eq:Rk} is then given by 
\begin{eqnarray}
A_k &=& \pi R_k^2 - \pi R_{k-1}^2\nonumber\\
&=&(18(k-1)-3)\pi r^2.  
\end{eqnarray}

Next, if we divide this remaining annulus of width $2r$ into $m$ equal parts radially, it is almost equivalent to (barring semicircles at the endpoints) discarding that many rectangles with widths of $r$ and heights of $2r$. Since the semicircles overlap with the discarded annuli, we are not underestimating the discarded region. This gives us the total discarded area within this annulus as
\begin{eqnarray}
    A_{k,\text{disc}}^\text{fry} = 2mr^2+(6(k-1)-3)\pi r^2,
\end{eqnarray}
where the second term is the discarded annulus due to the inner radius. To find the maximum valid $m$, we note that each segment must have an area of $\pi r^2$ which leads to
\begin{eqnarray}
    &A_k-A_{k,\text{disc}}^\text{fry}=m\pi r^2,\nonumber
    \\&\Rightarrow 12(k-1)\pi r^2 - 2mr^2=m\pi r^2,\nonumber
    \\&\Rightarrow m=\left\lfloor{\frac{12\pi(k-1)}{\pi+2}}\right\rfloor,\nonumber
    \\&\Rightarrow m\geq 7(k-1),
\end{eqnarray}
where $\lfloor.\rfloor$ is the floor function. This leads to a lower bound on the total segments by the fryum segmentation as $N = 3.5n^2 - 3.5n +1$ which is greater than the total hexagons of $N = 3n^2 - 3n +1$. This establishes that fryum segmentation packs more segments than the hexagonal close packing. 

\begin{figure}[t]
    \centering
    \includegraphics[width=0.9\linewidth, trim= 0 0.2cm 0 0.2cm, clip]{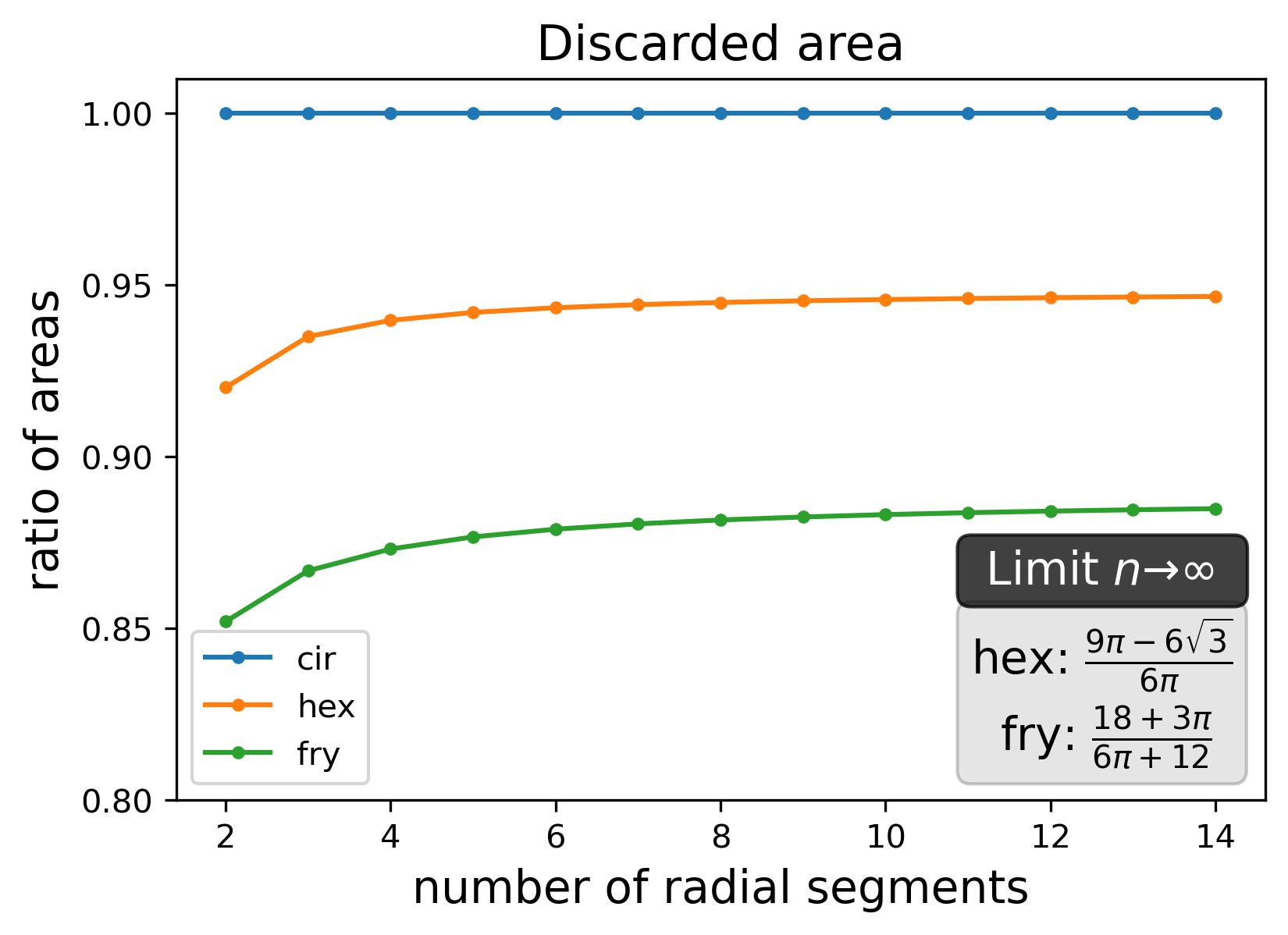}
    \caption{The plot shows the area discarded in each method as a fraction of the corresponding circle segmentation. Fryum always discards the least area.}
    \label{fig:discarded_limit}
\end{figure}

Next, the total discarded area in fryum segmentation is found by summing the discarded regions in each annulus as 
\begin{eqnarray}
    A_\text{disc}^\text{fry} &=& \sum_{k=2}^{n}A_{k,\text{disc}}^\text{fry} \nonumber
    \\&=& \left(\left\lfloor \frac{12}{\pi+2} n(n-1)\right\rfloor + 3(n-1)^2\right)\pi r^2
    \\&\leq& \left(5.33n^2 - 8.33n - 3\right)\pi r^2,
\end{eqnarray} 
where we have found the upper bound for the discarded region by ignoring the floor function. To compare the magnitudes of the discarded region, we plot them in comparison to the area discarded by the corresponding circular segmentation as seen in Fig.~\ref{fig:discarded_limit}. Clearly, we see that the fryum segmentation always wastes the least area.

Thus, we have shown that the fryum segmentation is the most optimal one. It packs the most segments while discarding the least area as compared to other regular uniform tilings. We stress the fact that the fryum wheel segmentation naturally extends itself to non-uniform probability distributions while retaining all its benefits like the maximal overlap of the discarded regions. The size of the atoms increases radially to maintain equiprobability. This causes non-maximal overlaps between the discarded regions of hexagons making them waste more area.

\section{Estimating the error due to border pixels for uniform distribution}\label{ap:border-error}
    In the following section, for simplicity, we assume the probability of detecting a photon over the detector area is uniform and that $\sigma_{2|1}$ is a square instead of a circle. Please note that the logic behind the subsequent calculations is still valid under these simplifying assumptions. 

\begin{figure}[ht]
        \centering
        \includegraphics[width=0.6\linewidth,trim= 0 0 0 0,clip]{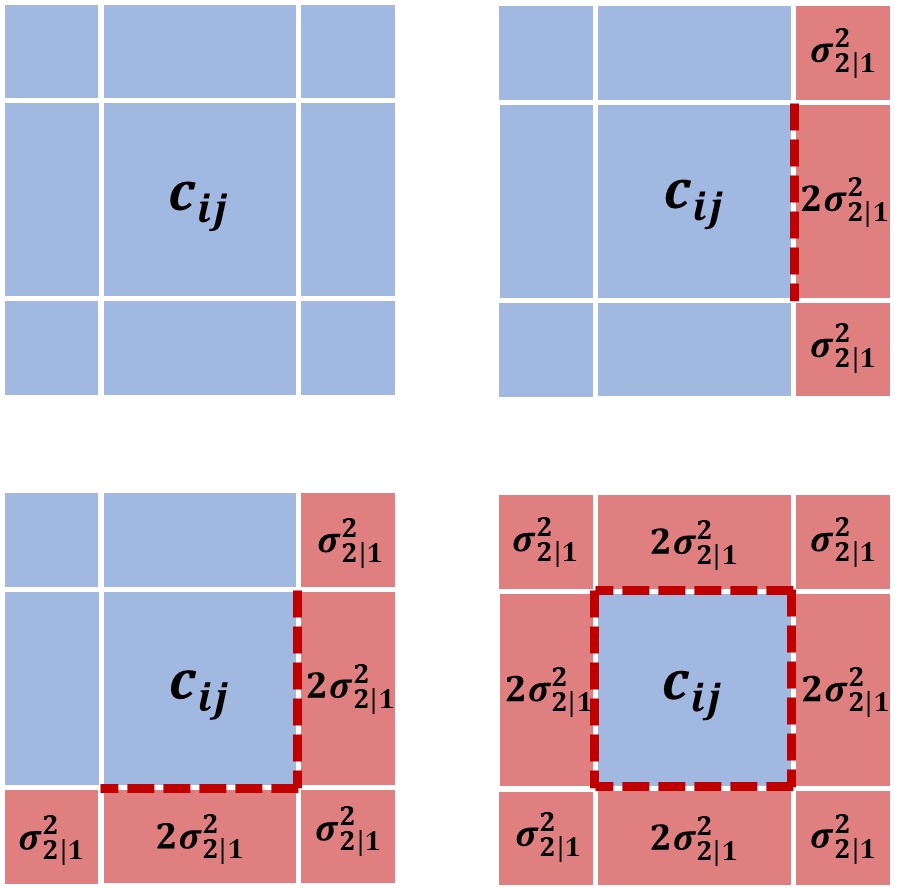}
        \caption{The four scenarios for the border pixels where blue and red colours represent two different macropixels on Alice's detector. The figure shows one full blue pixel $c_{ij}$ with 4 half pixels sharing an edge and 4 quarter pixels sharing a vertex with it. Detection of events in any of these pixels leads to a simultaneous event at $b_{ij}$ on Bob's detector. $c_{ij}$ has an area of $4\sigma_{2|1}^2$. The region in red contributes to the error in the central blue pixel. Starting from the top left, clockwise, the errors are: $\varepsilon=0,~\varepsilon=1/4,~\varepsilon=3/4,~\text{and }\varepsilon=7/16$.}
        \label{fig:border-pixels-counting}
    \end{figure}

    Consider all possible regions from Alice's detector that can lead to a simultaneous detection on Bob's pixel $b_{ij}$. As seen in Fig.~\ref{fig:border-pixels-counting}, it can come from nine possible locations, $c_{ij}$ and its 8 adjacent pixels. Pixels that share an edge contribute half as much as $c_{ij}$ and vertex ones share a quarter as much. Each quarter-square is $\sigma_{2|1}^2$ in area. 

    There are four cases of Alice's pixel $c_{ij}$ being adjacent to a pixel from another macropixel. As seen in Fig.~\ref{fig:border-pixels-counting} (and in Fig.~\ref{fig:discard_benefit}), the number of edges it shares with a pixel from another macropixel can be none (like $c_{22}$), one (like $c_{23}$), two (like $c_{33}$) or four (when each pixel is a macropixel). To calculate the error in correctly tagging Bob's pixel $b_{ij}$, we must first identify how many borders it shares with a neighbouring macropixel. If it shares no borders, $\varepsilon_{ij}=0$, if it shares one border, $\varepsilon_{ij}=1/4$, if it shares two borders, $\varepsilon_{ij}=7/17$ and if it shares four borders, $\varepsilon_{ij}=3/4$ (last case is only when each pixel itself is a macropixel). The total error in the whole detector would be $\varepsilon=\sum_{ij}\varepsilon_{ij}/N$ where $N$ is the total number of pixels. This error can then be used to estimate the key rate using Eq.~\eqref{eq:mod-keyrate}.

    In the example case considered in Sec.~\ref{subsec:discard-rationale}, out of the $36$ pixels ($20$ of which are border pixels), there are $16$ pixels that have one border and $4$ that have two borders. This gives us a total error of
    \begin{eqnarray*}
        \varepsilon=\frac{1}{36}\left(16\cdot\frac{1}{4} + 4\cdot\frac{7}{16}\right) \approx 0.16
    \end{eqnarray*}

    On the other hand, if we chose to maximize $d$ by treating each pixel as a macropixel, there would be $20$ pixels with two borders and $16$ pixels with four boundaries. This leads to an error of $\varepsilon\approx0.58$ which already is quite large. 

\end{document}